\documentclass[floatfix,showpacs,twocolumn]{revtex4}
\usepackage{amsmath, amsfonts, psfrag, epsfig}
\usepackage{rotating}

\newcommand{\hp}[1]{\hphantom{#1}}
\newcommand{\vp}[1]{\vphantom{#1}}
\newcommand{\E}{\mathbb E}
\newcommand{\st}{\tilde{\sigma}}
\newcommand{\nn}{\nonumber}

\newcommand{\JSP}{{J. Stat. Phys.} }
\newcommand{\JPC}{{J. Phys. C} }
\newcommand{\JPA}{{J. Phys. A} }
\newcommand{\PRE}{{Phys. Rev. E} }
\newcommand{\PRL}{{Phys. Rev. Lett.} }
\newcommand{\PRB}{{Phys. Rev. B} }
\newcommand{\PRA}{{Phys. Rev. A} }
\newcommand{\ZPB}{{Z. Phys. B} }
\newcommand{\PA}{{Physica A} }
\newcommand{\PTP}{{Prog. Theor. Phys.} }

\newcommand{\tst}{\textstyle}
\newcommand{\sst}{\scriptscriptstyle}
\newcommand{\pts}[1]{\mskip+#1\thinmuskip}
\newcommand{\nts}[1]{\mskip-#1\thinmuskip}

\newcommand{\N}{\mathbb{N}}
\newcommand{\R}{\mathbb{R}}

\newcommand{\eps}{\varepsilon}

\newcommand{\scc}{\mathcal S}
\newcommand{\ft}{\tilde{f}}

\newcommand{\asmax}{{{A'}_{z_l(y)}^{\sst (R)}}_{\mathrm{max}}}
\begin{document}
\title{Phase diagram of the random field Ising model on the Bethe lattice}

\author{Thomas Nowotny}
\affiliation{Institut f\"ur Theoretische Physik, Universit\"at Leipzig \\ 
P.O.B. 920, 04109 Leipzig, Germany\\}
\author{Heiko Patzlaff}
\affiliation{Institut f\"ur Theoretische Physik, Universit\"at Leipzig \\ 
P.O.B. 920, 04109 Leipzig, Germany\\}
\author{Ulrich Behn}
\affiliation{Institut f\"ur Theoretische Physik, Universit\"at Leipzig \\ 
P.O.B. 920, 04109 Leipzig, Germany\\}


\begin{abstract}
  The phase diagram of the random field Ising model on the Bethe lattice with
  a symmetric dichotomous random field is closely investigated with respect
  to the transition between the ferromagnetic and paramagnetic regime.
  Refining arguments of Bleher, Ruiz and Zagrebnov [{\em J.\ Stat.\ Phys.} {\bf
    93}, 33 (1998)] an exact upper bound for the existence of a unique
  paramagnetic phase is found which considerably improves the earlier
  results. Several numerical estimates of transition lines between a
  ferromagnetic and a paramagnetic regime are presented. The obtained results
  do not coincide with a lower bound for the onset of ferromagnetism proposed
  by Bruinsma [{\em Phys.\ Rev.\ B\/} {\bf 30}, 289 (1984)]. If the latter one
  proves correct this would hint to a region of coexistence of stable
  ferromagnetic phases and a stable paramagnetic phase.
\end{abstract}


\maketitle

\section{Introduction}
The random field Ising model (RFIM) has been studied extensively in theory
\cite{nattermann} as well as in experiment \cite{belanger}. The
one-dimensional model
\cite{BruinsmaAeppli}-\cite{behn6}
can be reformulated as a random iterated function system (RIFS) for an
effective field \cite{BruinsmaAeppli, rujan, gyoergyi1, brandtgross}. The
reformulation leads to an iteration of first order whereas standard transfer
matrix methods lead to iterated function systems of second order. This
considerable simplification allows deep insights into the effects of quenched
random fields on local thermodynamic quantities.

Being one-dimensional the Ising chain has no phase transitions for finite
temperature though. The RFIM on the Bethe lattice to the contrary exhibits
for not too high temperature at least a phase transition from ferromagnetic
behaviour for small random fields to paramagnetic behaviour for large fields
\cite{brz98, Bruinsma, Heiko}. The phase diagram is probably even much richer
\cite{Cieplak}. For $T=0$ hysteresis effects have been found and
investigated in detail \cite{shukla}.

The Bethe lattice (Cayley tree) is uniquely characterized by the two
properties that it is an infinite simple graph with constant vertex degree
and that it contains no loops. It is of order or degree $k$ if the vertex
degree is $k+1$. The Bethe lattice of degree $k=1$ is the one dimensional
lattice and the Bethe lattice of degree $k=2$ the well known binary tree.
Because the Bethe lattice contains no loops the RFIM on the Bethe lattice can
be reformulated to a (generalized) RIFS \cite{brz98, Bruinsma, brandt} for
the effective field like in the one-dimensional model \cite{rujan,
  brandtgross}. Therefore, the same powerful techniques as in the
one-dimensional case can be applied to gain insight into the mechanisms
driving the phase transition. Nevertheless the exact transition line in the
($T,h$) parameter plane is still not known. Recently, exact lower bounds for
the existence of a stable ferromagnetic phase as well as exact upper bounds
for the existence of a stable paramagnetic phase were proved \cite{brz98}. We
present an improved upper bound for the existence of a stable paramagnetic
phase based on this approach. These bounds are still far from the region
where the transition is expected though.  Therefore, we also develop several
criteria to detect the phase transition line numerically. It turns out that
the obtained results while being consistent with each other disagree
significantly with an early result by Bruinsma \cite{Bruinsma} who calculated
a lower bound for the onset of ferromagnetic behaviour. As Bruinsma's
argument rests on the differentiability of the density of the invariant
measure of the RIFS which was only proven for small $h$ and near $T_c$ there
are two possible interpretations. Either Bruinsma's bound is not true outside
the proven region of validity and the transition from ferromagnetic to
paramagnetic behaviour takes place at the smaller random field values found
in our numerical results or there is a region of coexistence of stable
ferromagnetic phases with a stable paramagnetic phase implying a phase
transition of first order in this region.

The paper is organized as follows. After introducing the model and our
notations in Sec.~\ref{secmodel} we present the improved exact upper bounds
for the onset of paramagnetism in Sec.~\ref{secexact}. In section
\ref{estimate} we give three criteria to estimate the transition line between
the ferromagnetic and paramagnetic regime. The expectation value of the local
magnetization is calculated directly and we extract an estimate for the
region of a stable ferromagnetic phase. We then study the average
contractivity of the RIFS of the effective field. This leads to an estimate
for the appearance of a stable paramagnetic phase for increasing random field
strength $h$. The third criterion is the independence of the effective field
from boundary conditions. It also provides an estimate for the stability
region of the paramagnetic phase. The implications of our results in
comparison to Bruinsma's approach are discussed in detail in the concluding
Sec.~\ref{secdiscuss}.

\section{The model} \label{secmodel}
\begin{figure}{
\psfrag{0}{$\scriptstyle y_0$}
\psfrag{1}{$\scriptstyle z_1(y)$}
\psfrag{2}{$\scriptstyle z_2(y)$}
\psfrag{3}{$\scriptstyle z_0$}
\psfrag{4}{$\scriptstyle y$}
\begin{center}\epsfig{width=0.75\columnwidth,file= 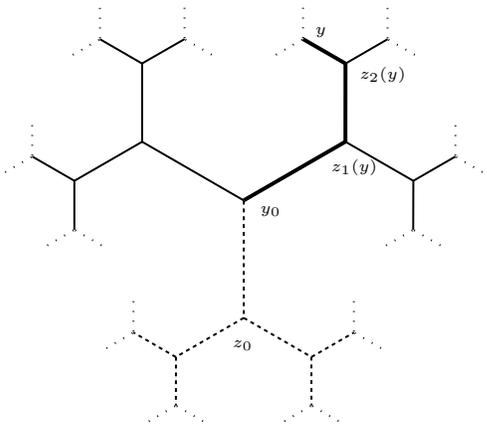}\end{center}}
\caption{Bethe lattice of degree $k=2$. The solid lines mark the part of the
  lattice denoted by $V^+$ and the dashed lines the part denoted by $V^-$.
  The roots of the two subtrees are denoted by $y_0$ and $z_0$ respectively.
  The thick line shows the unique path from a vertex $y \in \partial V_3$ at
  the boundary to the central vertex $y_0$ to illustrate the labeling along
  the path used in section \ref{secexact}.\label{fig1}}
\end{figure}


The formulation of the RFIM on a Bethe lattice requires some notations for
the underlying lattice. By $V$ we denote the set of vertices of the Bethe
lattice and $d(y, z)$ is the natural metric on the lattice given by the
length of the unique path connecting $y$ and $z$. Furthermore, $V_R := \{y
\in V : d(y,y_0) \leq R\}$ denotes the ball of radius $R$ around some
arbitrarily chosen central vertex $y_0$ and $\partial V_R:= \{y \in V: d(y,
y_0)= R\}$ its boundary, the sphere of radius $R$. In the following it will
be useful to decompose $V$ into two subtrees $V^+$ and $V^-$ with roots $y_0$
and $z_0$ in the way illustrated in Fig.~\ref{fig1}.

Introducing the
notation $\scc (y) := \{z \in \partial V_{R+1} : d(y,z)=1 \}$ for the
successors of $y \in \partial V_R$ the Hamiltonian of the RFIM on the Bethe
lattice reads
\begin{align}
  H_R(\{s_y\}_{y \in V_R}) = - \nts{4} \sum_{\substack{y \in V_{R-1} \\ z \in
      \scc(y)}} \nts{3} J s_y s_z - \nts{4} \sum_{y \in V_{R-1}} \nts{3} h_y
  s_y - \nts{4} \sum_{y \in \partial V_R} \nts{3} x_y^b s_y , \label{hamiltonian}
\end{align}
where $s_y$ denotes the classical spin at vertex $y$ taking values $\pm 1$, $J$
is the coupling strength, $h_y$ is the random field at site $y$ and $x_y^b$
the field at the boundary encoding the chosen boundary conditions. We
restrict ourselves to independent, identically distributed, symmetric
dichotomous random fields, i.\,e.,  $h_y= \pm h$ with probability $\frac{1}{2}$.
The canonical partition function
\begin{align}
  Z_R:= \nts{2} \sum_{\{s_y\}_{y \in V_R}} \nts{3} \exp (- \beta
  H_R(\{s_y\})) ,
  \label{partition}
\end{align}
where $\beta= (k_B T)^{-1}$ is the inverse temperature can be reformulated by a
method first introduced by Ruj\'an \cite{rujan} for the one dimensional
RFIM resulting in
\begin{align}
  Z_R= \nts{1} \sum_{s_{y_0} = \pm 1} \nts{3} \exp \beta \big( (x_{y_0}^{\sst
    (R)} + A(x_{z_0}^{\sst (R)})) \, s_{y_0} + {\tst \nts{5} \sum\limits_{y
      \in V_R \backslash \{y_0\}}} \nts{4} B(x_y^{\sst (R)}) \big) ,
  \label{partition2}
\end{align}
where the effective fields $x_z^{\sst (R)}$ are determined by the generalized
RIFS
\begin{align}
  x_y^{\sst (R)} = \sum\limits_{z \in \scc(y)} A(x_z^{\sst (R)}) + h_y 
  \label{iteration}
\end{align}
with boundary conditions $x_y^{\sst (R)} = x_y^b$ for $y \in \partial V_R$.
The functions $A$ and $B$ are given by
\begin{align}
  A(x) &= (2\beta)^{-1} \ln (\cosh \beta (x+J)/\cosh \beta (x-J)) ,  \\
  B(x) &= (2\beta)^{-1} \ln (4 \cosh \beta (x+J)\cosh \beta (x-J)).   
\end{align}
Note that the upper index ${}^{\sst (R)}$ of the effective field refers to
the radius of the sphere where the boundary conditions are fixed. The
partition function in the form (\ref{partition2}) is a partition function of
one spin $s_{y_0}$ in two effective fields $x_{y_0}^{\sst (R)}$ and
$A(x_{z_0}^{\sst (R)})$ which are both determined through the RIFS
(\ref{iteration}). The sum in (\ref{iteration}) implies that although $|A'| <
1$ for non-zero $T$, the RIFS is not necessarily contractive in contrast to
the one-dimensional case. A loss of contractivity indicates a phase
transition as is explained in more detail below.

Being functions of the random fields $h_y$ the effective fields are
random variables (RVs) on the random field probability space and the
iteration (\ref{iteration}) induces a Frobenius-Perron or Chapman-Kolmogorov
equation for their probability measure
\begin{align}
  \nu_y^{\sst (R)} (X) = \sum_{h_y = \pm h} \frac{1}{2} \Big(
  {\prod\limits_{z\in \scc(y)}}^{\nts{4} *} A_\# \nu_z^{\sst (R)}\Big)
  (X-h_y) ,
  \label{frobenius}
\end{align}
where ${\prod}^*$ denotes the convolution product of measures, $X$ is some
measurable set, $X- h_y := \{x-h_y | x \in X\}$ and $A_\#$ is the induced
mapping of $A$ on measures, i.\,e.,  $A_\# \mu (X) := \mu(A^{-1}(X))$. The
measures of the effective fields at the boundary are fixed by boundary
conditions, e.\,g.\ as $\nu_y^{\sst (R)} =\delta_{x_y^b}$, the Dirac measure at
$x_y^b$. Any other choice of the RVs $x^b_y$ is also possible though.

It was proved in \cite{brz98} that the existence of limiting Gibbs measures
with finite restrictions compatible with (\ref{hamiltonian}) and
(\ref{partition}), cf.\ \cite{Georgii}, implies the
weak convergence of the RVs $x_y^{\sst (R)}$, i.\,e.,  the weak convergence of
the measures $\nu_y^{\sst (R)}$ to measures $\nu_y$ in the limit $R \to
\infty$. For homogeneous boundary conditions $x_y^b \equiv x^b$ for all $y
\in V$ the measures $\nu_y$ are all identical and will be
denoted by $\nu$.

Before we can present our results on phase transitions in the RFIM on the
Bethe lattice some more properties of the RIFS (\ref{iteration}) and the
function $A$ are necessary.  $A(x)$ is a monotonic function in $x$. For a
given random field configuration $\{h_y\}_{y \in V_R^+}= \{\sigma_y h\}_{y
  \in V_R^+}$, $\sigma_y = \pm$, we denote the composite function mapping the
effective fields in $\partial V_{R+1}^+$ to the effective field at $y_0$ by
$f_{\{\sigma\}_R}$. Here, $\{\sigma\}_R$ is the tree of $k^{R+1}-1$ symbols $\pm$
characterizing the configuration of the random field and $k$ is the degree
of the Bethe lattice. These composite functions are monotonic in the sense
that if $x_y^b \geq {x'}_y^b$ for all $y \in \partial V^+_{R+1}$ then
$f_{\{\sigma\}_R} (\{x_y^b\}) \geq f_{\{\sigma\}_R}(\{{x'}_y^b\})$. In the
same way they are monotonic with respect to the random field,
$f_{\{\sigma\}_R} (\{x_y^b\}) \geq f_{\{\sigma'\}_R}(\{x_y^b\})$ if $\sigma_y
\geq \sigma'_y$ for all $y \in V_R^+$.  Furthermore, there exists an invariant
interval $I=[x^*_-, x^*_+]$ with the property that if $x_y \in I$ for all $y
\in \partial V_{R+1}^+$ then also $f_{\{\sigma\}_R}(\{x_y\}) \in I$ for any
random field configuration $\{\sigma\}_R$. Here, $x^*_-$ and $x^*_+$ are the
fixed points of the composite functions for homogeneous $\{-\}$ and homogeneous
$\{+\}$ configuration of the random field respectively. Since $A(x) = -A(-x)$,
these fixed points are symmetric, $x^*_- = - x^*_+$.

\section{Upper bounds for the existence of a unique paramagnetic phase}
\label{secexact}
In this section we present an exact upper bound for the existence of a unique
paramagnetic phase in terms of the random field strength $h$. This bound
improves earlier results in \cite{brz98}.

Throughout this section we will use effective fields $g_y := A(x_y)$ in close
analogy to the notation in \cite{brz98}. This has some advantages in the
calculation which will become clear below. The iteration (\ref{iteration})
for $g_y$ reads
\begin{align}
  g_y^{\sst (R)} = \left\{ \begin{array}{ll}
      g_y^b & (\mbox{for $y \in \partial V_R$}) \\[0.2cm]
      A \big( \nts{5} \sum\limits_{\pts{4} z \in \scc(y)} \nts{5} g_z^{\sst
        (R)} + h_y ) & (\mbox{otherwise}) 
    \end{array} \right. ,  \label{iteration2}
\end{align}
and we denote the composite functions mapping the effective fields
$\vp{\}_y}\smash{\{g_y\}_{y \in \partial V_{R+1}^+}}$ to $g_{y_0}$ by
$\tilde{f}_{\{\sigma\}_R}$. They have the same monotonicity properties as the
composite functions $f_{\{\sigma\}_R}$.

In order to prove the existence of a unique pa\-ra\-mag\-netic phase it is
sufficient to show that the RVs $g_y$ do not depend on the boundary
conditions $\{g_y^b\}$ in the limit $R \to \infty$ for any choice of the
boundary conditions.  We use the notation $g_y^+$ for the effective field at
$y \in V$ for homogeneous boundary conditions
$g_y^b \equiv g^*_+$ in the limit $R \to \infty$ and $g_y^-$ for the effective field resulting from the
corresponding negative boundary conditions $g_y^b \equiv g^*_-$ where $g^*_+
=A (x^*_+)$ and $g^*_- = A(x^*_-)$. For $g^+_{y_0}$ and $g^-_{y_0}$ we use
the shorthand notations $g^+$ and $g^-$. Note that the dependence of the
effective fields on the random field configurations is suppressed in this
notation.

Inspired by the proof for the existence of a unique paramagnetic phase for
the RFIM on the Bethe lattice of degree $2$ for almost all random field
configurations and $2 < h < 3$ in \cite{brz98}, we
investigate the expectation value
\begin{align}
\E_{\{\sigma\}} (|g^+ -g^- |) . \label{eqn17}
\end{align}
The monotonicity of the composite functions $\tilde{f}_{\{\sigma\}_R}$
implies that if this expectation value is zero for the two extremal boundary
conditions chosen above then it is zero for any two sets of boundary
conditions. This then implies that the RV $g_{y_0}$ is independent of the
boundary conditions for almost all
random field configurations. The goal of this section is therefore to find a
criterion for the random field strength $h$ which implies that the
expectation value (\ref{eqn17}) is zero.  Because of the monotonicity of the
composite functions $\ft_{\{\sigma\}_R}$ we have $g^+ \geq g^-$ and thus
$|g^+ - g^-| = g^+ - g^-$. Therefore, we consider
\begin{align}
  \E_{\{\sigma\}} \big( g^+-g^-) &= \int \! d\eta(\{\sigma\}) \,
  (g^+ - g^-) \\
  &= \sum_{\{\sigma\}_R} \nts{19} \int\limits_{\pts{25}\{\st\}_R = \{\sigma\}_R}
  \nts{21} d\eta(\{\st\}) \, \big(g^+(\{\st\}) - g^-(\{\st\})\big) , \nn 
\end{align}
where $\eta$ is the product measure of the probability measures of the random
fields $h_y = \sigma_y h$. In the second step the integration was split up
into a sum of a finite number of integrals over sets of configurations with
fixed symbols $\{\sigma\}_R$ in $V_R$ and arbitrary $\{\st\} \in V \backslash
V_R$. Using the recursion relation
(\ref{iteration}) the integrand can be expressed as a function of the effective
fields $\{g_y^+\}_{y \in \partial V_R}$ on the boundary of $V_R$,
\begin{align}
  & g^+(\{\st\})-g^-(\{\st\}) \nn \\
  & \hp{g} = \ft_{\{\st\}_{R-1}} (\{g_y^+\}_{y \in \partial
  V_R}) - \ft_{\{\st\}_{R-1}}(\{g_y^-\}_{y \in \partial V_R}) \nn \\
  & \hp{g} = \sum_{y \in \partial V_R}  \partial_{g_y} \ft_{\{\st\}_{R-1}}
  (\{\delta_z\}_{z \in \partial V_R}\}) \cdot (g_y^+ - g_y^-) .
  \label{expandeq}
\end{align}
In the second step the mean value theorem has been used for $\ft_{\{\st\}_R}$
and $\delta_z \in [g_z^-, g_z^+]$ are appropriately chosen. The partial
derivatives in (\ref{expandeq}) are bounded from above by $\prod_{l= 0}^{R-1}
\asmax$, where $\asmax$ is an upper bound on the maximum of $A'(x)$ for $x
\in [g_{z_l(y)}^-, g_{z_l(y)}^+]$, the interval of possible values of the
effective field at the vertices $z_l(y)$ along the unique path from $y$ to
$y_0$, cf Appendix \ref{appa} for details. This bound only depends on
$\{\tilde{\sigma}\}_R=\{\sigma\}_R$ and hence is independent of the
integration. Thus,
\begin{align}
  & \E_{\{\sigma\}}(g^+ - g^-)  \label{eqnr} \\
  & \hp{\E} \leq \nts{1} \sum_{\{\sigma\}_R \pts{1}}
  \nts{1} \sum_{\pts{1} y \in \partial V_R} \prod_{\pts{1} l=0}^{R-1}
  \asmax \nts{6} \int\limits_{\{\st\}_R = \{\sigma\}_R} \nts{9}
  d\eta(\{\st\}) \, (g_y^+ - g_y^-) . \nn
\end{align}
The remaining integral for each $y$ is bounded from above by $2^{-|V_R|} \,
\E_R$ where $\E_R = \max_{y \in \partial V_R} \E_{\{\sigma\}} ( g_y^+ - g_y^-
)$, cf Appendix \ref{appb}. We therefore obtain
\begin{align}
  \E_0 &= \E_{\{\sigma\}}(g^+ - g^- ) \nn \\
  &\leq \nts{0} \sum_{\{\sigma\}_R} 2^{-|V_R|} \, \sum_{y \in \partial V_R} \;
  \prod_{l= 0}^{R-1} \, \asmax \, \E_R . \label{eqn121}
\end{align}
The finite sums commute and as $\asmax$ is obtained with homogeneous boundary
conditions the sums $\sum_{\{\sigma\}_R}$ are identical for all $y \in
\partial V_R$ such that the sum over $y$ can be replaced by a factor
$|\partial V_R|= k^R$ yielding
\begin{align}
  \E_0 \leq K \, \E_R ,
\end{align}
where
\begin{align}
  K \, :=  \sum_{\{\sigma\}_R} \nts{0}
  2^{-|V_R|} \, k^R \, \prod_{l= 0}^{R-1} \,
  \asmax . \label{eqn5} 
\end{align}
Because of the translation invariance of the Bethe lattice these
considerations can be applied recursively. This implies $\E_0 \leq K^r \E_{r
  \cdot R}$. If the factor $K$ is less than $1$ for any parameters ($T$, $h$)
we immediately obtain $\E_0=\E_{\{\sigma\}}(|g^+ - g^-|) = 0$ as $\E_{r\cdot
  R}$ is uniformly bounded by $2 g^*_+$ for all $r \in \N$ and therefore $K^r
\E_{r\cdot R} \to 0$ for $r \to \infty$. By translation symmetry this result
holds for all $g_y$ with $y \in V$. As $|g^+ - g^-| \geq 0$ the vanishing
expectation even implies $|g^+ - g^-| = 0$ for almost all realizations
$\{\sigma\}$ of the random field.

The reason for using $g_y$ instead of $x_y$ is now easily explained. If we
used the effective fields $x_y$ instead of $g_y$ the product over derivatives
of $A$ would be from $l=1$ up to $R$. This gives a less precise estimate
because $x_y$ with $y \in \partial V_R$ is less restricted than $x_{y_0}$ and
therefore the bound for $A'(x_y)$ with $y \in \partial V_R$ is greater than
the one for $A'(x_{y_0})$.

\begin{figure}[t]
\psfrag{h}{\turnbox{180}{\footnotesize $\nts{20}$ random field $h$}}
\psfrag{T}{\footnotesize $\nts{20}$ temperature $k_B T$}
\psfrag{Tc}{$\nts{4} \scriptstyle k_B T_c$}
\psfrag{0}{$\scriptstyle 0$}
\psfrag{1}{$\scriptstyle 1$}
\psfrag{2}{$\scriptstyle 2$}
\psfrag{3}{$\scriptstyle 3$}
\centerline{\epsfig{file=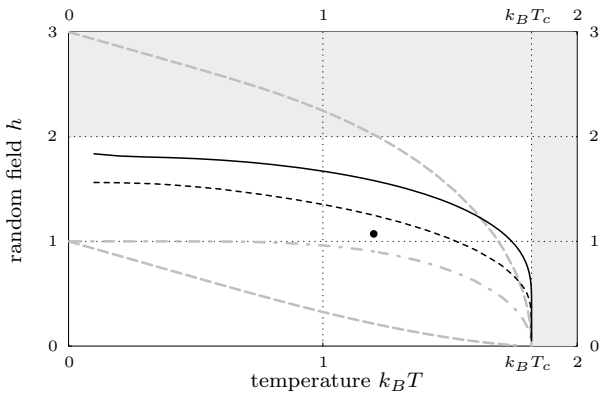, width= 0.9\columnwidth}}
\caption{Exact upper bound for the existence of a stable paramagnetic phase on
  the Bethe lattice of degree $k=2$ (solid line). The bound was obtained as
  described in the text with all random field configurations at $R=4$. The
  dashed line is a similar upper bound obtained by considering a sample
  of $10^4$ realizations of the random field at $R=11$ using
  the complete sum (\ref{eqn121}). Close to $T=0$ the problem is numerically
  unstable; results are presented only for $T \geq 0.1$. The
  large dot was obtained for $R=23$ using (\ref{eqn5}) and $10^5$ random
  field configurations. In the shaded region the result
  of \cite{brz98} for the existence of a unique paramagnetic phase applies.
  The grey dashed lines are the ferromagnetic and the antiferromagnetic lines
  \cite{Bruinsma}, cf.\ also \cite{brz98}, and the grey dash dotted line is
  Bruinsma's lower bound for the existence of a stable ferromagnetic phase
  \cite{Bruinsma}. ($J=1$) \label{fig2}}
\end{figure}

 To apply the criterion obtained above we evaluated $K$ on a
computer. The calculation time is proportional to the number of random field
configurations on $V_R$ and thus asymptotically grows for, e.\,g., $k=2$ as
$2^{2^{R}}$. Therefore, the calculation was restricted to $R \leq 4$ (for $R=
5$ each data point in an array of $20 \times 40$ points would take about 3
days on a Pentium II 350MHz). The solid line in Fig.~\ref{fig2} shows the
upper bound for the existence of a unique paramagnetic phase obtained for $R=
4$.

To estimate the results for $R > 4$ we relied on statistical methods and
sampled random field configurations. Doing so it is saving time not to
exploit the symmetry and use (\ref{eqn121}) instead of (\ref{eqn5}). The
resulting bound for $R= 11$ and a sample of $10^4$ random field
configurations is the dashed line in Fig.~\ref{fig2}.

As the obtained bound systematically depends on the radius $R$ it is tempting
to try to extrapolate to $R = \infty$.  For $T=1.2$ we obtained an extremely
good fit for the data sampled at $R=5$, \ldots, $23$ using $h_c(R) = a + b
R^c$. However, the result was $a= h_c(\infty) \approx 0$ which is not
realistic. Other fits, e.\,g.~omitting data for small $R$ or using different
functional forms, yielded values between $-0.9$ and $0.45$. We suspect that a
na$\ddot{\mbox{\i}}$ve choice of the functional form of $h_c(R)$ does not allow
realistic extrapolation results for these bounds in the case of the Bethe
lattice.

\section{Estimates for the transition line}
\label{estimate}
\subsection{Direct calculation of the magnetization} \label{secmagnet}

Even though the bounds presented in the preceding section
considerably improve former analytical results they are still far away from
the region where the phase transition from paramagnetic to ferromagnetic
behaviour is suspected. In \cite{Bruinsma} Bruinsma claimed to have found a
lower bound in $h$ for the existence of a ferromagnetic phase which is in
the relevant parameter region, cf.\ Fig.~\ref{fig2}. To check this bound
and to get a good numerical approximation of the transition line we developed
several numerical criteria for the existence of a ferromagnetic phase or the
existence of a stable paramagnetic phase.

The most obvious criterion for the
existence of a ferromagnetic phase is a non-vanishing expectation value for
the magnetization for small but non-zero boundary conditions. The expectation
value for the local magnetization at the spin in the center is given by
\begin{align}
 m &:= \E_{\{\sigma\}} \langle s_{y_0} \rangle \nn \\
  &= \int d\nu(x) \, d\nu(y) \, \tanh \big( \beta (x + 
  g(y)) \big)  \label{eqn15}
\end{align}
where $\langle \cdot \rangle$ denotes the thermodynamic average,
$\E_{\{\sigma\}}$ the expectation value with respect to all random field
configurations and $\nu$ is the limit measure of the effective field for
homogeneous boundary conditions $x^b_y \equiv x^b$ for all $y \in V$ in the
limit $R \to \infty$. To
approximate $\nu$ we generated a large number of random field configurations
on a finite region $V_R$ and calculated the corresponding effective field
$x_{y_0}^{\sst (R)}$. The obtained values were then sorted into small boxes
of length $\eps$. The resulting histogram was used as an approximation of
$\nu_{y_0}(b_i)=: \nu_i$ where $b_i$ is the $i$-th box. Explicitly this
yields for the magnetization
\begin{align}
  m &\approx \sum_{i}\sum_{j} \, \nu_i \nu_j \,
  \tanh \beta (x_i + g(y_j)) .
\end{align} 
where the points $x_i$ and $y_j$ were chosen as the center of box $i$ and $j$
respectively.

\begin{figure}[t]
\psfrag{0}{\raisebox{-0.3ex}{$\scriptstyle 0$}}
\psfrag{a}{$\nts{2} \scriptstyle 0.5$}
\psfrag{b}{$\scriptstyle 1$}
\psfrag{1}{\raisebox{-0.3ex}{$\scriptstyle 1$}}
\psfrag{2}{\raisebox{-0.3ex}{$\scriptstyle 2$}}
\psfrag{3}{\raisebox{-0.3ex}{$\scriptstyle 3$}}
\psfrag{0.5}{\raisebox{-0.3ex}{$\scriptstyle 0.5$}}
\psfrag{T}{$\nts{20}$ \raisebox{-0.5em}{\footnotesize temperature $k_B T$}}
\psfrag{h}{\turnbox{180}{\footnotesize $\nts{15}$ random field $h$}}
\psfrag{m}{$m$}
\psfrag{Tc}{\raisebox{-0.3ex}{$\nts{4} \scriptstyle k_B T_c$}}
\centerline{\epsfig{file=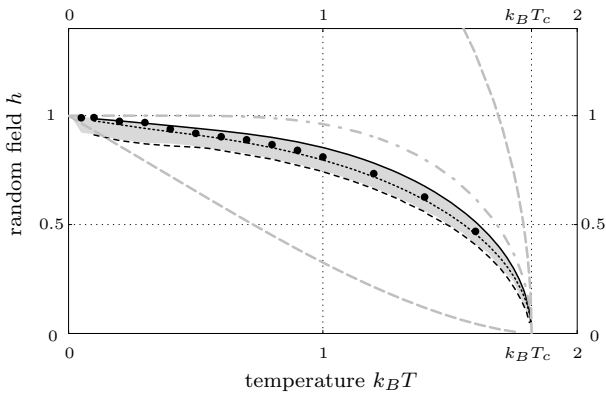, width= 0.9\columnwidth}}
\caption{
  Estimates for the transition line between between para- and ferromagnetism.
  The dashed line separates regions where the average magnetization decreases
  or increases with the distance from the boundary for homogeneous boundary
  conditions $x^b \equiv 0.01$ obtained by comparing the configurational
  averages over $4\cdot 10^5$ up to $64 \cdot 10^5$ samples at distances $9$
  and $13$ from the boundary. The dotted line gives the boundary of the
  region in which the iteration (\ref{iteration}) is contractive on the
  average above and non-contractive below. It was obtained by sampling $10^5$
  field configurations and evaluating (\ref{avg2}) with $R= 13$ and $R_1= 4$.
  The solid line was obtained using (\ref{eqn10}) at $R=4$. The big dots were
  obtained by evaluating (\ref{eqn117}) for $R=20$ and between $10^5$ and
  $2.4 \cdot 10^6$ random field configurations. The grey lines are as in
  Fig.\ \ref{fig2}. The grey shaded region marks the region all our numerical
  results (including considerably more than shown explicitly here) are
  contained in.  ($k=2$, $J=1$) \label{fig3}}
\end{figure}


Assuming that the magnetization in the center varies monotonically with the
radius $R$ of the finite volume $V_R$ one would expect to observe a
monotonically increasing magnetization in the ferromagnetic regime and a
monotonically decreasing magnetization in the paramagnetic regime for
increasing $R$.  Therefore, the dashed contour in Fig.~\ref{fig3} which
divides the regions in which the numerical estimate of the magnetization
is increasing or decreasing with increasing $R$ is a good estimate for the
transition line. This type of estimates only slightly depends on the chosen
boundary condition and iteration depth but the obtained results significantly
disagree with Bruinsma's bound.

\subsection{Average contractivity of the RIFS} \label{seccontract}
For zero boundary conditions there is a paramagnetic state for any
temperature $T$ and random field strength $h$. The stability of this state is
tied to the average contractivity of the RIFS (\ref{iteration}). If it is
globally contracting the paramagnetic state is stable and unique. If it is at
least contracting on the average for some interval around zero, the
paramagnetic phase is stable but the existence of other stable phases is not
a priori excluded. The investigation of the contractivity of the RIFS was
first proposed in \cite{Heiko}.

We generated a set $\Sigma_R$ of random field configurations $\{\sigma\}_R$
on a finite ball $V_R$ and calculated the image of a small initial interval
$I^b =[-x^b, x^b]$. Because of the monotonicity of $f_{\{\sigma\}_R}$ the
image of this interval at vertex $y$ is $I_y=[x_{y}^{\sst (R)} (-x^b),
x_{y}^{\sst (R)} (x^b)]$. To estimate the average contractivity of the RIFS
we compared the average length $1/k^{R_1} \sum_{y \in \partial V_{R_1}}
|I_y|$ at the vertices $y \in \partial V_{R_1}$ to the length $|I_{y_0}|$ at
the central vertex $y_0$. As the effective fields at all $y \in \partial
V_{R_1}$ contribute to the effective field at $y_0$ we consider the average
interval lengths at vertices in $\partial V_{R_1}$ instead of individual values.
To minimize the influence of the somewhat arbitrary choice of the initial
interval the comparison was performed for $R_1 \ll R$.

There are two ways of performing the comparison.  Either one first averages
over the lengths $|I_y|$ at all $y \in \partial V_{R_1}$ then calculates the
quotient of $|I_{y_0}|$ and this average length in $\partial V_{R_1}$ and
average over the sample $\Sigma_R$ of random field configurations at the end,
\begin{align}
  \bigg\langle \frac{|I_{y_0}|}{\frac{1}{k^{R_1}} \sum_{y \in \partial
      V_{R_1}} |I_{y}|} \bigg\rangle_{\Sigma_R} .  \label{avg2}
\end{align}
Or one first averages $|I_y|$ over all $y \in \partial V_{R_1}$ and all
random field configurations as well as $|I_{y_0}|$ over the same random field
configurations and calculates the quotient at the end, 
\begin{align}
  \frac{\big\langle |I_{y_0}| \big\rangle_{\Sigma_R}}
  {\big\langle \frac{1}{k^{R_1}} \sum_{y \in \partial V_{R_1}} |I_{y}|
    \big\rangle_{\Sigma_R}} . \label{avg3}
\end{align}
The two averaging procedures (\ref{avg2}) and (\ref{avg3}) yield identical
results and thus obviously are equivalent.

If the images of the initial interval contract on the average for a finite
iteration of (\ref{iteration}) we expect complete contraction to length zero
for infinite iteration. This corresponds to a stable paramagnetic phase.
Therefore, the contour in the ($T,h$) parameter plane at which the average
quotient of band lengths switches from greater than $1$ below to less than
$1$ above is an estimate for the stability region of the paramagnetic phase.
The resulting estimated transition line is shown for $R=13$, $R_1=4$, the
initial interval $I^b=[-0.01, 0.01]$ and $10^5$ random field configurations
as the dotted line in Fig.~\ref{fig3}. Again, the agreement of the obtained
results for various boundary conditions and iteration depths is satisfactory
but there is a large deviation from Bruinsma's line.

\subsection{Independence of the effective fields from boundary conditions} \label{slopeprodsec}
A related criterion for the existence of a stable paramagnetic
phase is the independence of the effective field from boundary conditions. As
in Sec.~\ref{secexact} we use the effective fields $g_y^{\sst (R)}$ rather
than $x_y^{\sst (R)}$. We consider boundary conditions $\{g^b_y\}_{y \in
  \partial V_R}$ taking values in a small interval $[-g^b, g^b]$. Through the
iteration with (\ref{iteration}) the effective fields $g_y^{\sst (R)}$ are
functions of the boundary conditions
\begin{align}
  g_y^{\sst (R)} = \ft_{\{\sigma\}_{R-n-1}(y)}
  (\{g^b_z\})
\end{align}
where the function $\ft_{\{\sigma\}_{R-n-1} (y)}$ has $k^{R-n}$ arguments for
$y \in \partial V_n$ and it is the identity if $R = n$.  For simplicity
and without loss of generality we restrict the following discussion to
$g_{y_0}^{\sst (R)}$. The boundary conditions can be written as $g^b_y =
g^b \, \hat{g}^b_y$ where $\hat{g}^b_y$ takes values in $[-1,1]$. To
investigate how the effective fields depends on the boundary conditions
we consider the expectation value of the derivative of
$g_{y_0}^{\sst (R)}$ with respect to $g^b$, the {\em strength} of the applied
boundary condition
\begin{align}
  & \E_{\{\sigma\}_{R-1}} \Big| \frac{d}{dg^b} g_{y_0}^{\sst (R)} \Big| \nn \\
  & \hp{\E} \leq \E_{\{\sigma\}_{R-1}} \sum_{y \in \partial V_R}
  \partial_{g_y} \ft_{\{\sigma\}_{R-1}} (\{g^b_z\}_{z \in \partial V_R})
  \, |\hat{g}^b_y|   \nn \\
  & \hp{\E} \leq \sum_{y \in \partial V_R} \E_{\{\sigma\}_{R-1}}
  \prod_{l=0}^{R-1} A'(x_{z_l(y)}^{\sst (R)}(x^b))
\end{align}
where $x_{z_l(y)}^{\sst (R)}(x^b)$ denotes the effective fields along the
unique path from $y$ to $y_0$ with homogeneous
boundary conditions $x^b_y \equiv A^{-1}(g^b)$. Now one can estimate similar as
in Appendix \ref{appa}  
\begin{align}
  & A'(x_{z_l(y)}^{\sst (R)}(x^b)) \nn \\
  & \hp{A} \leq \min \{ A' (\max\{ x_{z_l(y)}^{\sst
    (R)}(x^b), 0\}, A'(\min \{x_{z_l(y)}^{\sst (R)}(-x^b),0\}\} \nn \\
  & \hp{A} =: \asmax (x^b) . \label{eqn21}
\end{align}
As the boundary conditions are homogeneous this implies
\begin{align}
  \E_{\{\sigma\}_{R-1}} \Big| \frac{d}{dg^b} g_{y_0}^{\sst (R)} \Big| \leq
  k^R \E_{\{\sigma\}_{R-1}} \prod_{l= 0}^{R-1} \asmax (x^b) . \label{eqn10}
\end{align} 
If the right hand side vanishes for $R \to \infty$ the effective field
$g_{y_0}$ is on the average independent from boundary conditions taking
values in $[-g^b, g^b]$.  By determining the parameter region in which the
right hand side of (\ref{eqn10}) vanishes for $R \to \infty$ we therefore get
an upper bound on the emergence of a stable paramagnetic phase.
As our calculations are limited to finite $R$, convergence
to zero is assumed if the obtained values of (\ref{eqn10}) for $R > 0$ are
smaller than $1$ which is the value for $R= 0$.

For the Bethe lattice of degree $k= 2$, radius $R = 4$ and $g^b= 0.01$ the
right hand side of (\ref{eqn10}) was evaluated.  The contour between values
smaller than $1$ above and greater than $1$ below is shown as the solid line
in Fig.~\ref{fig3}. For $R > 4$ we again relied on sampling random field
configurations instead. The resulting transition lines for various iteration
depths and boundary conditions are comparable to the results of the preceding
two sections and are all contained in the grey region in Fig. \ref{fig3}.

If we consider the derivative of the effective field at $y_0$ in
the case of boundary conditions $g^b \equiv 0$ we have
\begin{align}
  \E_{\{\sigma\}_{R-1}}  \frac{d g_{y_0}^{\sst (R)}}{dg^b} (\{0\})  
  &=  \E_{\{\sigma\}_{R-1}} \sum_{y \in \partial V_R} \prod_{l=0}^{R-1}
  A'(x_{z_l(y)}^{\sst (R)}(\{0\})) . \label{eqn117} 
\end{align}
If this derivative does not tend to zero for some
parameters $(T,h)$ and $R \to \infty$ there is no stable
paramagnetic phase. By determination of the parameter region in which this is
the case we get a lower bound on the emergence of a stable paramagnetic
phase. The numerical results are the large dots in Fig.~\ref{fig3}.

\section{Discussion} \label{secdiscuss}
In order to interpret the discrepancies between Bruinsma's result and our
numerical investigations we briefly review Bruinsma's argument
\cite{Bruinsma} in our language. \begin{figure}
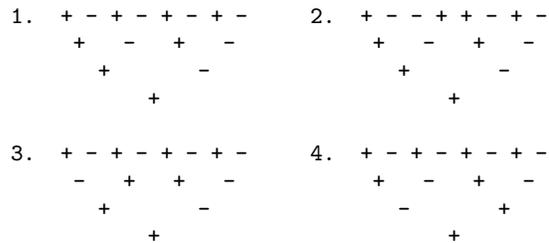

\begin{center}
  \begin{minipage}{0.9\columnwidth}
\begin{verbatim}
1.  + - + - + - + -     2.  + - - + + - + - 
     +   -   +   -           +   -   +   -  
       +       -               +       -     
           +                       +        
\end{verbatim}
\begin{verbatim}
3.  + - + - + - + -     4.  + - + - + - + -
     -   +   +   -           +   -   +   -
       +       -               -       +
           +                       +
\end{verbatim}
  \end{minipage}
\end{center}
\caption{Four equivalent chessboard configurations at $R=4$. The second
  configuration is obtained from the first by permutation of two subtrees of
  a vertex in the sphere $\partial V_2$, the third one by permutation of two
  subtrees of a vertex in the sphere $\partial V_1$ and the last one by
  permutation of the two subtrees of the root itself. \label{fig4}}
\end{figure}
%
 The probability measures
$\nu_y$ of the effective fields $x_y$ are fixed points of the
Frobenius-Perron equation, Eq.~(\ref{frobenius}).  They can be approximated
by finite iterations of some initial probability densities (boundary
conditions) $\nu^b_y$ for $y \in \partial V_R$. If the support of $\nu^b_y$
is a subset of the invariant interval $I$, the support of $\nu_{y_0}$ is a
subset of the images of $I$ by functions $f_{\{\sigma\}_R}$. These images are
called {\em bands}. The left and right boundary of the bands are the
effective fields corresponding to homogeneous boundary conditions $x^b_y
\equiv x^*_+$ and $x^b_y \equiv x^*_-$ for $y \in \partial V_R$, respectively.
The investigation of the structure of the set of bands has proved to be a
powerful tool in the treatment of the one-dimensional random field Ising
model
\cite{gyoergyi1,brandtgross,behn1,behn2,bene1,bene2,tanaka1,behn5,behn6}. In
contrast to the one-dimensional case the bands are highly degenerate here,
i.\,e., different configurations of the random field result in the same band.
This is due to the invariance of the model with respect to permutations of
subtrees for homogeneous boundary conditions.  The most degenerate bands
correspond to the two {\em chess-board} configurations, cf.\ Fig.~\ref{fig4},
of the random field with $+h$ or $-h$ at $y_0$ respectively. There are
$2^{2^{R-1}-1}$ equivalent random field configurations in the case of the
Bethe lattice of degree $k=2$ and radius $R$. As the total number of
configurations is $N= 2^{2^R -1}$ the most degenerate bands have a weight of
$2^{2^{R-1}-1} / 2^{2^R -1} = 2^{-2^{R-1}} \sim 1/ \sqrt{N}$. The bands with
the least weight are the bands corresponding to homogeneous $+h$ or $-h$
random field configurations.  They have the weight $1/N$. The weights of all
other bands are distributed between these values.

Bruinsma used boundary conditions $\nu^b_y \equiv \delta_{x^b}$ for some $x^b
\in \R$ and iterated (\ref{frobenius}). He only considered the lowest
and highest weight terms corresponding to the least and the most degenerate
bands. The highest weight terms obey a recursion relation. The fixed points
of this relation can be calculated and it is straightforward to determine for
which temperatures $T$ and random field strengths $h$ they are symmetric to
the origin. Proving differentiability of the density $\rho$ of the invariant
measure $\nu$ in a neighbourhood of $T= T_c$ and $h=0$, Bruinsma concluded
that an asymmetric position corresponds to asymmetric maxima of $\rho$ of
non-zero weight and therefore to the existence of a ferromagnetic phase.

The symmetric position corresponds to complete
contraction of the most degenerate bands such that the asymmetric boundary
condition has no effect in the limit of infinite iteration.  The asymmetric
position on the other hand occurs if the most degenerate bands do not
completely contract. Seen this way the argument
above is our criterion of average contractivity of the RIFS in
Sec.~\ref{seccontract} except that it is restricted to the contractivity
of one specific band instead of the average contractivity.

\begin{figure}
\psfrag{a}{\raisebox{-1mm}{$\scriptstyle \nts{3} -2$}}
\psfrag{c}{\raisebox{-1mm}{$\scriptstyle \nts{1} 0$}}
\psfrag{e}{\raisebox{-1mm}{$\scriptstyle \nts{1} 2$}}
\psfrag{f}{$\scriptstyle \nts{3} 0$}
\psfrag{g}{} 
\psfrag{h}{$\scriptstyle \nts{7} 0.5$}
\psfrag{xl}{\raisebox{-1.8ex}{$\nts{2} x$}}
\psfrag{yl}{\raisebox{1.3ex}{$\nts{5} \rho(x)$}}
\parbox{0.45 \columnwidth}{\raggedright \footnotesize {\bf a)} $h= 0.3$} \hfill
\parbox{0.45 \columnwidth}{\raggedright \footnotesize {\bf b)} $h= 0.961159$} \hfill
\epsfig{file=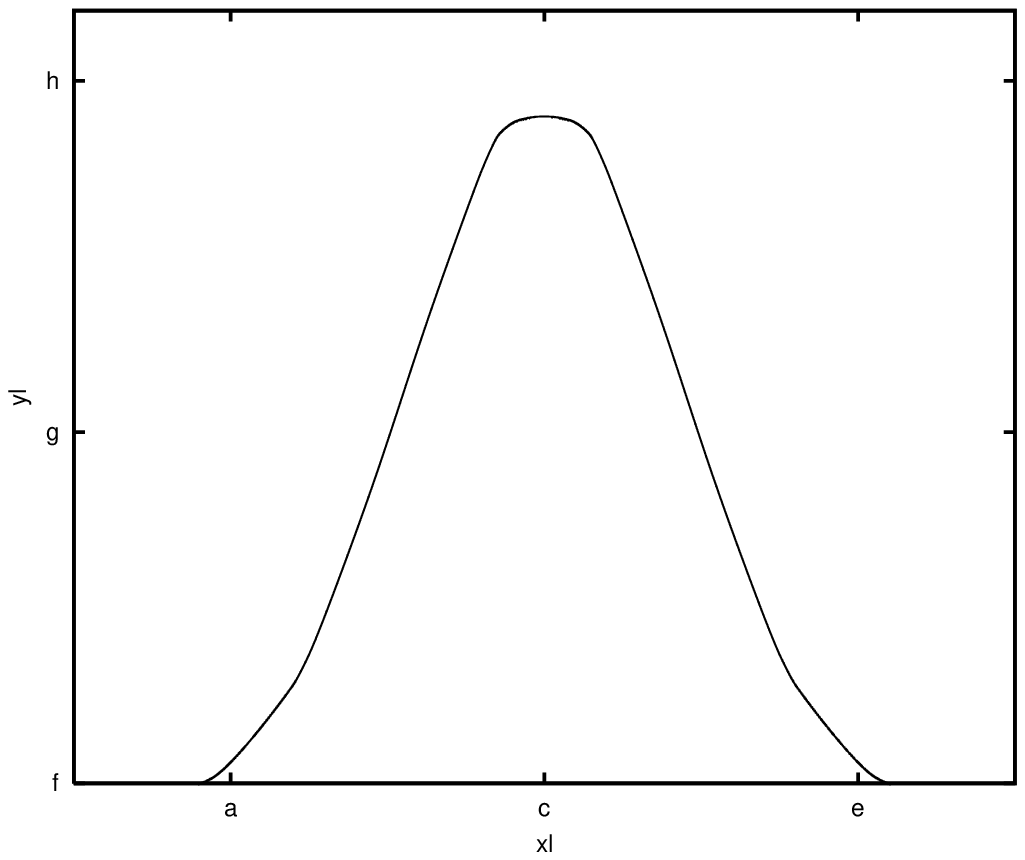, width= 0.45\columnwidth} \hfill
\epsfig{file=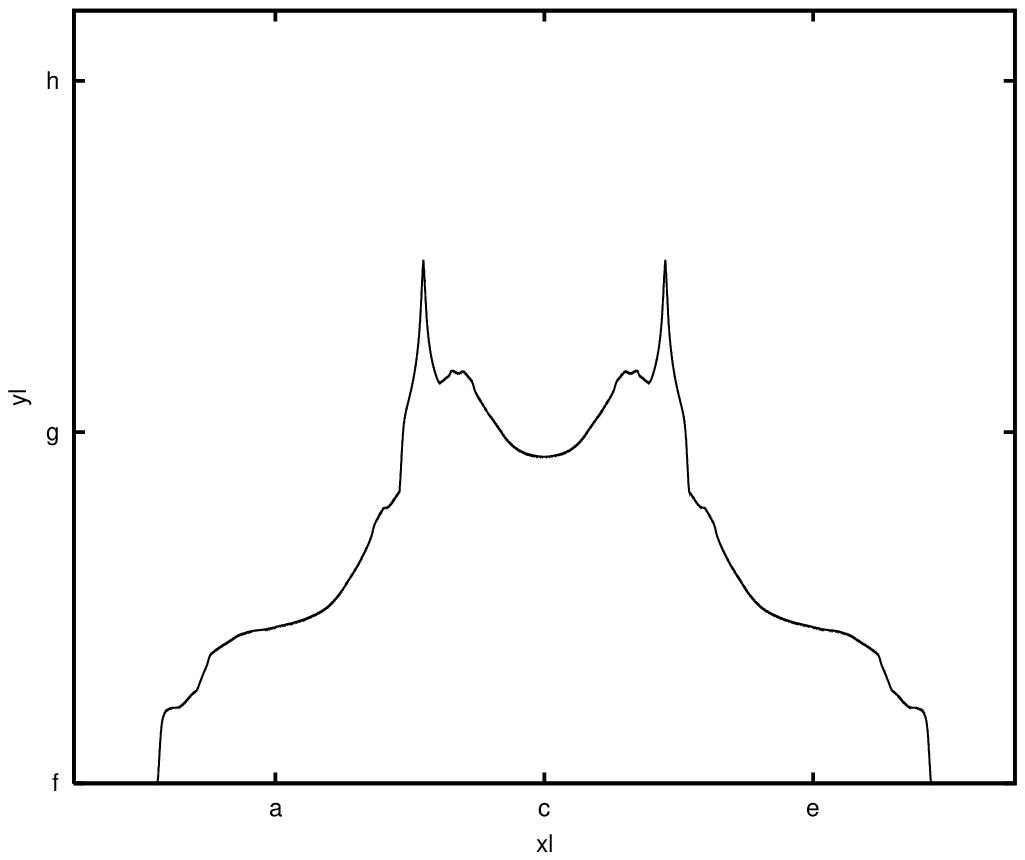, width= 0.45\columnwidth}
\caption{Approximations of the invariant measure density $\rho$ obtained
  from four-fold application of (\ref{frobenius}) to the equipartition $\rho^b
  = |I|^{-1} \cdot {\boldsymbol 1}_I$ on the invariant interval $I$. There
  are no maxima at $\pm h$ in a) whereas the two maxima at $\pm h$ in b)
  are already so pronounced that differentiability of $\rho$ is
  questionable. The random field strength in b) was chosen such that the point
  $(T,h)$ is very close to Bruinsma's bound. ($k=2$, $\beta=J=1$) \label{fig5}}
\end{figure}
%
 There are two problematic points in the reasoning above.
Firstly, it is not clear whether the location of local maxima in a
differentiable measure density really is given by the most degenerate bands.
For small $h$ this actually seems not to be the case, cf.\ Fig.~\ref{fig5}a.
As the maxima are at $\pm h$ and therefore close to zero for small $h$ it is
difficult to argue based on numerical data though. The example in
Fig.~\ref{fig5}b shows however that the maxima {\em are} present for
sufficiently large $h$.

Secondly, the differentiability of the invariant measure density has been
proved only in a neighbourhood of $T= T_c$ and $h=0$ whereas for large $h$ or
small $T$, the measure density $\rho$ is clearly {\em not} differentiable. It
is unclear whether it is differentiable in the region of the lower bound,
cf.\ Fig.~\ref{fig5}b.

The disagreement of our numerical work with Bruins\-ma's lower bound
therefore allows two interpretations. Either Bruinsma's bound is not true
outside the proven region of validity because the most degenerate bands are
not a sufficient indicator for the symmetry of $\rho$ when the measure
density is not differentiable.  Or in the region between our upper bounds for
the existence of a stable paramagnetic phase and Bruinsma's lower bound for
the onset of ferromagnetism a stable paramagnetic phase coexists with the ---
also stable --- ferromagnetic phases. This would imply the existence of a
first order phase transition and hysteresis loops depending on the strength
of the random field in contrast to the hysteresis at $T=0$ \cite{shukla}
which depends on the homogeneous offset of the random field.

In this paper we improved exact upper bounds for the existence of a unique
paramagnetic phase in Sec.~\ref{secexact} which is a further step towards the
exact determination of the phase diagram of the RFIM on the Bethe lattice.
Furthermore, we presented numerical work leading to various estimates for the
actual phase transition line. The direct calculation of the expectation value
of the local magnetization in Sec.~\ref{secmagnet}, the investigation of the
average contractivity of the RIFS (\ref{iteration}) at large iteration depths
in Sec.~\ref{seccontract} and the numerical calculation of the derivative of
the effective field with respect to the strength of the boundary condition in
section \ref{slopeprodsec} provided estimates for the stability region of the
paramagnetic phase. All results are in good agreement while all disagreeing
with the earlier result of Bruinsma \cite{Bruinsma}. This disagreement
motivates further investigations whether the bound for the onset of
ferromagnetism given in \cite{Bruinsma} needs to be reconsidered or whether
there really is a coexistence region for stable ferromagnetic and a stable
paramagnetic phase.

The work was partially supported by the Cusanuswerk and the DFG
(Graduiertenkolleg ``Quan\-ten\-feld\-theo\-rie'').

\begin{appendix} 
\section{}
\subsection{Bound on the partial derivatives in (\ref{expandeq})}
\label{appa}
The partial derivatives in (\ref{expandeq}) are given by
\begin{align}
  & \partial_{g_y} \ft_{\{\st\}_{R-1}}  (\{\delta_z\}_{z \in
    \partial V_R}\}) \nn \\
  & \hp{\partial_{g_y}} = \prod_{l= 0}^{R-1}
  A'\big(f_{\{\st\}_{R-1-l}(z_l(y))} (\{\eps_z\}_{z \in \partial V_R})\big) ,
    \label{eqn7} 
\end{align}
where $\eps_z = A^{-1}(\delta_z)$, $z_l(y) \in \partial V_l$ are the vertices
along the unique path from $y$ to $y_0$, cf.\ also Fig.~\ref{fig1}, and
$\{\st\}_{R-1-l}(z_l(y))$ are the signs of the random field configuration on
the subtree of depth $R-1-l$ with root $z_l(y)$. The terms
$f_{\{\st\}_{R-1-l}(z_l(y))} (\{\eps_z\}_{z \in \partial V_R})$ are effective
fields $x_{z_l(y)}^{\sst (R)}$ corresponding to boundary conditions $\{x_z^b
= \eps_z\}_{z \in \partial V_R}$. We write $x_{z_l(y)}^{\sst (R)}(\eps)$ for
these fields and $x_{z_l(y)}^{\sst (R)}(x^+)$ and $x_{z_l(y)}^{\sst
  (R)}(x^-)$ for the corresponding effective fields with boundary conditions
$x_z^b = x_z^+$ and $x_z^b = x_z^-$ for $z \in \partial V_R$. We then can
estimate
\begin{align}
  & A'\big( x_{z_l(y)}^{\sst (R)}(\eps)\big) \nn \\
  & \hp{A'} \leq \max\big\{ A'\big(x_{z_l(y)}^{\sst (R)}(\eps')\big) \,
  \big| \, \eps'_z \in [x_z^-,x_z^+] , z \in \partial V_R \big\} \nn \\
  & \hp{A'} = \max\big\{A'(x) \, \big| \, x\in [ x_{z_l(y)}^{\sst (R)}(x^-),
  x_{z_l(y)}^{\sst (R)}(x^+)]\big\} \nn \\
  & \hp{A'} = \min\big\{ A'(\max\{ x_{z_l(y)}^{\sst (R)}(x^-) , 0\}), \nn \\
  & \hp{ \hp{A'} = \min\big\{ } A'(\min\{ x_{z_l(y)}^{\sst (R)}(x^+) , 0\})
  \big\} . \label{theeqn}
\end{align}
In the last step we used that the maximum of $A'$ in an interval $[a,b]$ is
at $a$ if $a \geq 0$, at $b$ if $b \leq 0$ and at zero in all other cases.  As
the effective fields can never be larger than $x^*_+$ and never smaller than
$x^*_-$ we can for $z \in \partial V_R$ estimate $x_z^- \geq h_z + k A(x^*_-)
= x_z^{\sst (R+1)} (x^*_-)$ and $x_z^+ \leq h_z + k A(x^*_+) = x_z^{\sst
  (R+1)} (x^*_+)$. This allows to replace $x^+$ and $x^-$ in the argument
of $x_{z_l(y)}^{\sst (R)}$ in (\ref{theeqn}) and with $x_{z_l(y)}^{\sst
  (R)}(x_y^{\sst (R+1)}(x^*_\pm))= x_{z_l(y)}^{\sst (R+1)}(x^*_\pm)$ we get
\begin{align}
  A'\big( x_{z_l(y)}^{\sst (R)}(\eps)\big) & \leq \min\big\{ A'(\max\{
  x_{z_l(y)}^{\sst (R+1)}(x^*_-) , 0\}), \nn \\
  & \hp{\leq \min\big\{ } A'(\min\{ x_{z_l(y)}^{\sst
    (R+1)}(x^*_+) , 0\}) \big\} \nn \\
  & =: \asmax . \label{eqn20}
\end{align} 
Inserting (\ref{eqn20}) into (\ref{eqn7}) then yields
\begin{align}
\partial_{g_y} \ft_{\{\st\}_{R-1}} (\{\delta_z\}_{z \in
    \partial V_R}\})  \leq  \prod_{l= 0}^{R-1} \asmax .
\end{align}

\subsection{Bounds on the integrals in (\ref{eqnr})}
\label{appb}
Using the independence of the RVs $g_y$ of
the signs $\{\sigma_z\}_{z \in V_R \backslash \{y\}}$ and denoting the
number of vertices in $V_R$ by $|V_R|$, i.\,e.,  $|V_R|= (k^{R+1}-1)/(k-1)$,
one obtains
\begin{align}
  & \int\limits_{\{\st\}_R = \{\sigma\}_R} \nts{9} d\eta(\{\st\})
  \, (g_y^+ - g_y^-) \nn \\
  & \hp{\int\limits_{\{\st\}_R = \{\sigma\}_R} \nts{9} } = 2^{-|V_R|+1}
  \nts{2} \int\limits_{\st_y = \sigma_y}
  \nts{4} d\eta(\{\st\}) \, (g_y^+ - g_y^-) \nn \\
  & \hp{\int\limits_{\{\st\}_R = \{\sigma\}_R} \nts{9} } = 2^{-|V_R|} \,
  \E_{\{\sigma\}} ( g_y^+ - g_y^- \, | \, \st_y = \sigma_y) .
\end{align}
The function $A$ is antisymmetric which implies $g_y^+(\{-\sigma\}) = -
g_y^-(\{\sigma\})$ and therefore
\begin{align}
  \E ( g_y^+ - g_y^- \, | \, \sigma_{y} = +) =\E (g_y^+ -g_y^- \, | \,
  \sigma_y= -) ,
\end{align}
implying
\begin{align}
  & \E(g_y^+ -g_y^-) \nn \\
  & \hp{\E} = \tst \frac{1}{2} \E(g_y^+ -g_y^- \, | \, \sigma_y = +) +
  \frac{1}{2} \E(g_y^+ -g_y^- \, | \, \sigma_y = -) \nn \\
  & \hp{\E} = \E (g_y^+ -g_y^- \, | \, \sigma_y=
  \sigma) ,
\end{align}
for any $\sigma \in \{-,+\}$. Setting
\begin{align}
  \E_R &:= \max_{y \in \partial V_R} \E_{\{\sigma\}} ( g_y^+ - g_y^- \, | \,
  \st_y = \sigma_y) \nn \\ 
  &= \max_{y \in \partial V_R} \E_{\{\sigma\}} ( g_y^+ - g_y^- ) ,
\end{align}
this finally yields
\begin{align}
  \int\limits_{\{\st\}_R = \{\sigma\}_R} \nts{9}
  d\eta(\{\st\}) \, (g_y^+ - g_y^-) \leq 2^{-|V_R|} \, \E_R .
\end{align}
\end{appendix}


\end{document}